\documentclass[twoside,12pt]{article}
\usepackage{graphicx,color}
\def\Journal#1#2#3#4{{#1} {#2} (#4) #3}

\def\NPA{{\em Nucl. Phys.} A}

\def\NPB{{\em Nucl. Phys.} B}

\def\PLB{{\em Phys. Lett.} B}

\def\PRL{\em Phys. Rev. Lett.}

\def\PREP{\em Phys. Rep.}

\def\PRD{{\em Phys. Rev.} D}
\def\PRC{{\em Phys. Rev.} C}

\def\RMP{{\em Rev. Mod. Phys.}}

\newcommand{\be}{\begin{equation}}
\newcommand{\ee}{\end{equation}}
\newcommand{\bea}{\begin{eqnarray}}
\newcommand{\eea}{\end{eqnarray}}

\newcommand\Real{\Re e}
\newcommand\Imag{\Im m}

\newcommand{\Hept}{$^4$He($\vec{e},e^\prime \vec{p}$)}
\newcommand{\Hpt}{$^1$H($\vec{e},e^\prime\vec{p}$)}

\newcommand{\Ee}{\ensuremath{E_e}} 
\newcommand{\thetae}{\ensuremath{\theta_e}}

\newcommand{\GEpGMp}{\ensuremath{G_E^p/G_M^p}}
\newcommand{\GEn}{\ensuremath{G_E^{n}}}

\newcommand{\GMn}{\ensuremath{G_M^{n}{}}}
\newcommand{\Q}{\ensuremath{Q^{2}{}}}
\newcommand{\updeg}{$^{o}$}

\begin{document}
\title{\vspace{1cm} 
Recent Experimental Results from JLab}
\author{
Kees\ de Jager \\
\\
Jefferson Laboratory, Newport News, VA 23606, USA}
\maketitle
\baselineskip=11.6pt
\begin{abstract}
Preliminary results of a selected few recent experiments at JLab are presented.
\end{abstract}
\baselineskip=14pt
\section{Introduction}

The JLab 6~GeV research program in hadronic physics can be separated  
into three broad areas of investigation:  the structure of the nuclear building 
blocks; the structure of nuclei; and symmetry tests in nuclear physics. Here, the results of eight recent experiments are presented that answer detailed questions of two of these areas:
\begin{itemize}
\item{measure precisely the nucleon's charge and magnetization distribution}
\item{determine the internal structure of the nucleon in the valence region}
\item{develop the experimental methods for performing tomography of the nucleon}
\item{probe the nuclear interior with a controlled impurity to learn about deeply-lying shell structure}
\item{clarify the short-range nature of nucleon-nucleon interactions in nuclei and compare the properties of bound nucleons with free ones}
\item{test chiral perturbation theory by studying the properties of Goldstone bosons}
\end{itemize}

At present the Continuous Electron Beam Accelerator Facility (CEBAF) accelerates electrons to 6 GeV by recirculating the beam 
four times through two superconducting linacs, each producing an energy 
gain of 600 MeV per pass. In the past decade there has been impressive progress at JLab in improving the quality of the polarized beam, such that a beam can be delivered routinely with a polarization of $\sim$85\% and an intensity of up to 200 $\mu$A. 
The base instrumentation in Hall A has been used for experiments which require high luminosity and high 
resolution in momentum and/or angle of at least one of the reaction 
products. The central elements are the two High Resolution 
Spectrometers (HRS), to which recently a third spectrometer has been added with a large acceptance (BigBite). 
The CEBAF Large Acceptance Spectrometer (CLAS) 
in Hall B is used for experiments that require the detection of 
several, loosely correlated particles in the hadronic final state 
at a limited luminosity. The Hall C facility has generally been used for experiments which 
require high luminosity at moderate resolution. The core 
spectrometers are the High Momentum Spectrometer (HMS) and the Short 
Orbit Spectrometer (SOS). The HMS has a maximum momentum of 7.6 GeV/$c$.
Of the experiments presented here, only  \Hept\ used just the base equipment, all others needed additional specially designed instrumentation and three experiments, Primakoff, hypernuclear spectroscopy in Hall C and \GEn\, used only the beam line in the Hall it was run.

An upgrade of the CEBAF accelerator to 12 GeV has recently obtained the CD-2 approval, an essential component of the formal DOE process for large projects. The project also includes the construction of a fourth experimental hall and the upgrade of two of the existing halls. It is on track to provide a first 11 GeV beam in 2013.

\section{The Structure of the Neutron}

Although the partonic structure of the proton has been studied in great detail, much less is known about that of the neutron, mainly due to the unavailability of a free neutron target. Thus, the information about the neutron had to be extracted from scattering data on nucleons bound in nuclei which resulted in substantial theoretical uncertainties. A novel detection technique, called BONUS - for Barely Off-shell NUcleon Structure - has been recently implemented in the CLAS detector in Hall B that allows to measure the inclusive electron scattering off an almost free neutron, eliminating to a large extent nuclear binding effects. This technique uses a low-momentum recoil detector to tag slow backward-moving spectator protons - with  a momentum as small as 70 MeV/$c$ - in coincidence with the scattered electron in the reaction $^2$H($e,e^\prime p_s$)X, semi-inclusive deep-inelastic scattering (SIDIS).

\begin{figure}[h!]
\begin{center}
\begin{minipage}[t]{12 cm}
\includegraphics[width=0.8\linewidth]{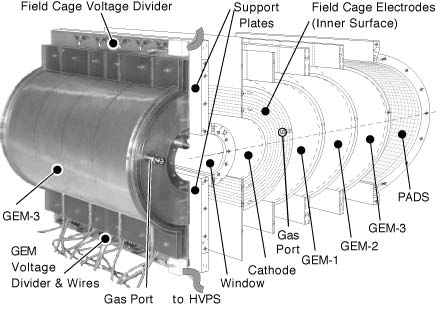}
\end{minipage}
\begin{minipage}[t]{16.5 cm}
\caption{Partially exploded schematic diagram of the BONUS detector, showing the three cylindrical GEM detectors, the cathode surface and the pad read-out. \label{fig:bonuslayout}}
\end{minipage}
\end{center}
\end{figure}

The structure functions of the nucleon reflect a variety of aspects of QCD, from asymptotic freedom at large momenta down to confinement at the hadronic scale. From measurements of these structure functions in the scaling region one can extract the parton distribution functions (PDF), the information on the momentum and spin carried by the quarks as a function of Bjorken $x$. In the non-perturbative region the structure functions in the nucleon resonance regime can be compared to QCD-inspired models. The applicability of quark-hadron duality can be probed by comparing the resonance structure and the partonic structure. Although the high-quality JLab data on $F_2^p$ and $F_1^p$ in the resonance region, when averaged over the resonances, show remarkable agreement with the extrapolation of DIS results in the perturbative region down to lower $Q^2$ at comparable $x$, no such comparison has been possible for the neutron. Another open question of similar impact is the behavior of the structure functions as $x$ approaches 1, in the region where the $u$ and $d$ valence quarks dominate, when all of the nucleon momentum is carried by a single quark. Simple phenomenological models like SU(6) symmetry predict a significantly different behavior than perturbative QCD or quark models with improved hyperfine interactions.

Figure \ref{fig:bonuslayout} shows a schematic lay-out of the detector. It basically consists of a radial time-projection chamber (RTPC) surrounding a thin deuterium target. The RTPC is relatively small, 20 cm long and 13 cm in diameter, situated at the standard CLAS target position, inside the superconducting solenoid designed for the DVCS program. The target is a 20 cm long, 6 mm diameter kapton cylinder, holding deuterium gas at 7.5 atm. The 30 cm long solenoid is used to suppress the background due to M{\o}ller electrons in both the recoil detector and the forward drift chambers. The RTPC is designed to provide position and timing information sufficient to identify low-momentum protons in the back hemisphere. It consists of 3 cylindrical Gas Electron Multiplier (GEM) detectors - the first implementation of curved GEMs - mounted around a kapton inner window with read-out pads on the outside. Charged particles leave a trail of electron-ion pairs on passing through the sensitive gas inside the RTPC. A radial electric field forces the electrons to drift outwards while the field focusing due to the holes in the GEMs causes avalanche multiplication. Finally, the signal is collected on the outer surface by individual pads. The location of the pads provides position information and the arrival time provides a measure of the radius where the electrons were produced. From the amount of charge produced the $dE/dx$ can be estimated and thus the mass of the particle traversing the RTPC.

\begin{figure}[h]
\begin{center}
\begin{minipage}[t]{12 cm}
\includegraphics[width=1.0\linewidth]{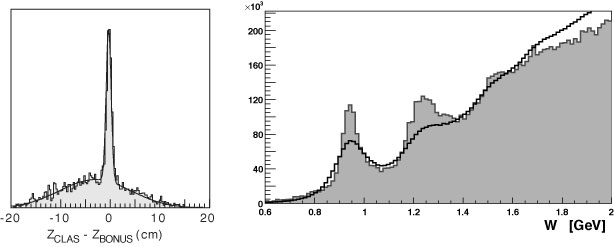}
\end{minipage}
\begin{minipage}[t]{16.5 cm}
\caption{Left: distribution of the difference in the vertex reconstruction using either the information from the CLAS drift chambers or that from the BONUS detector for coincident events detected during the commissioning run. Right: invariant mass $W$ spectrum reconstructed using the backward-going spectator protons detected in BONUS (grey area), compared to a $W$ spectrum from DIS off the deuteron.\label{fig:bonuscomm}}
\end{minipage}
\end{center}
\end{figure}

The measurement of tagged structure functions in SIDIS off the deuteron, by detecting a slow recoil proton in the back hemisphere, will resolve the ambiguities introduced by nuclear effects in extracting neutron structure functions from DIS off the deuteron. By detecting the spectator proton with the smallest momentum feasible, one minimizes the degree to which the struck neutron is off-shell and thus the associated uncertainties. Also, at such low momenta the deuteron wave function (or more accurately the spectral function) is well known. In the back hemisphere corrections due to target fragmentation are predicted to be very small. Lastly, final-state interaction (FSI) effects have been calculated to contribute less than 5\% to the spectral function - and thus to the extraction of $F_2^n$ - for spectator momenta less than 100 MeV/$c$ at angles larger than 130\updeg.
The BONUS detector was successfully installed and commissioned in the fall of 2005, followed by a first two-month production run for experiment E03-012\cite{e03-012} with a 40 nA beam  at an energy of 3.2 GeV. The BONUS detector operated to full satisfaction at a luminosity of $\sim10^{33}$ cm$^{-2}$s$^{-1}$. The vertex could be reconstructed with an accuracy of $\sim$1 cm (see fig. \ref{fig:bonuscomm} (left)). In the right part of that figure the invariant-mass spectrum reconstructed with the BONUS detector (grey area) is compared to one obtained from DIS off the deuteron. Note that the width of the resonances has decreased thanks to the absence of Fermi motion effects in the BONUS data. The data obtained will provide accurate values of the ratio $F_2^n/F_2^p$, and thus of $d(x)/u(x)$, up to an $x$-value of $\sim$0.5.

\section{Nuclear Medium Modifications}

Possible modifications of the properties of nucleons by the nuclear medium that they are bound in, have long been the subject of study. Early conclusions drawn from the longitudinal/transverse ratio of the cross section for electron-induced proton knock-out have since been convincingly refuted\cite{lapikas}. However, several calculations suggest a measurable deviation of the nucleon electro-magnetic form factors (EMFF) from their free-space values already at modest momentum-transfer values $Q^2 \leq$ 2.5 GeV$^2$. Polarization transfer in quasi-elastic nucleon knock-out, ($\vec{e}$, e$^\prime \vec{N}$), provides the required sensitivity to the properties of the nucleon in the nuclear medium. In free electron-nucleon scattering the measured ratio of the transverse and longitudinal polarization, transferred to the proton, is directly proportional to the ratio of the charge and magnetic form factors (see section \ref{GEn}). This property has been used in a study of polarization transfer in $^4$He, chosen as a target nucleus because of its high density and relatively simple structure, that facilitates microscopic calculations.

\begin{figure}[h]
\begin{center}
\begin{minipage}[t]{8 cm}
\includegraphics[width=1.5\linewidth]{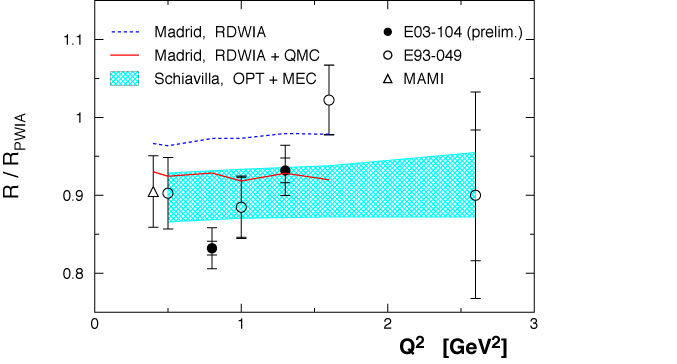}
\end{minipage}
\begin{minipage}[t]{16.5 cm}
\caption{Super-ratio $R/R_{PWIA}$ as a function of $Q^2$ from Mainz\cite{mainz} and E93-049\cite{strauch} (open symbols) along with preliminary results from E03-104 (filled circles): $R$ is the ratio of transverse to longitudinal polarization of the knocked-out proton in 
\Hept\ compared to the same ratio for \Hpt. The baseline $R_{PWIA}$ is the value of $R$ obtained in a plane-wave calculation, to account for "trivial" effects of free $vs.$ moving protons. The data are compared to calculations from the Madrid group\cite{udias} and Schiavilla {\it et al.}\cite{schiavil}.\label{fig:poltrans}}
\end{minipage}
\end{center}
\end{figure}

Earlier \Hept\ polarization-transfer experiments were performed at Mainz\cite{mainz} at $Q^2$ = 0.4 GeV$^2$ and in Hall A\cite{strauch} at $Q^2$ = 0.5, 1.0, 1.6 and 2.6 GeV$^2$. A recent experiment, again in Hall A\cite{e03-104}, has extended these measurements in 2006 with high precision at 0.8 and 1.3 GeV$^2$. The data were taken in quasi-elastic kinematics at low missing momentum in order to minimize FSI and meson-exchange corrections. One of the two high-resolution spectrometers in Hall A was used to detect the scattered electron, the other to detect the knocked-out proton. The azimuthal distribution of protons scattered by a graphite analyzer was measured in a focal-plane polarimeter\cite{fpp} to yield the polarization observables. The two-body character of the reaction was identified with the missing-mass technique. Polarization transfer in the \Hpt\ reaction was measured at the same kinematics in order to study the differences between the in-medium polarization transfers and the free values. The proton form-factor ratio \GEpGMp\ extracted from the hydrogen data was found to be in excellent agreement with previous polarization-transfer data\cite{gayou}. The results of the present experiment are expressed in terms of the polarization-transfer double ratio, in which the $^4$He polarization-transfer ratio is normalized to the $^1$H one, measured under identical conditions:

\begin{equation}
R = \frac{(P_x^\prime/P_z^\prime)_{^4\rm{He}}}{(P_x^\prime/P_z^\prime)_{^1\rm{H}}}
\end{equation}

Nearly all systematic uncertainties cancel in $R$, the remaining small uncertainties are predominantly due to the spin transport in the HRS spectrometer. The induced proton polarization $P_y$ was also extracted from the helicity-independent FPP data and is a direct measure of FSI effects.

The preliminary results for the double ratio $R$ are shown in fig. \ref{fig:poltrans} and are clearly in agreement with the existing data. The inner bars indicate the statistical errors, the outer ones the linear sum of statistical and systematic errors. The final systematic errors are expected to be significantly smaller than those shown here. The relativistic distorted-wave impulse-approximation (RDWIA) calculations from Udias {\it et al.}\cite{udias} by themselves overpredict the data by $\sim$6\%. On the other hand, if the density-dependent medium-modified form factors calculated by Lu {\it et al.}\cite{lu} are folded into Udias' calculations (QMC), a good agreement is obtained with most data, indicating possible modifications of the proton EMFF by the nuclear medium. However, the observed suppression of the super-ratio has been equally well described by a more traditional calculation  by Schiavilla {\it et al.}\cite{schiavil} with free form factors in which charge-exchange FSIs and meson-exchange and isobar current effects have been included. The high statistics of the present experiment\cite{e03-104} will allow a study of the individual polarization observables as a function of the missing momentum $p_m$. It should be pointed out that the observed low value of $R$ at $Q^2$ = 0.8 GeV$^2$ could be interpreted as an indication of a $Q^2$-dependence of $R$ not predicted by any of the calculations shown.

\begin{figure}[h]
\begin{center}
\begin{minipage}[t]{8 cm}
\includegraphics[width=1.4\linewidth]{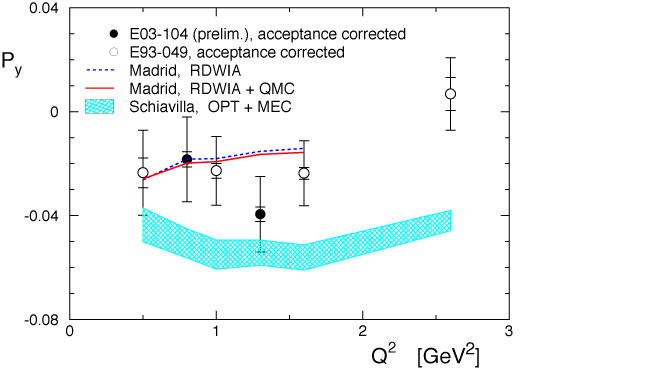}
\end{minipage}
\begin{minipage}[t]{16.5 cm}
\caption{Induced polarization data from Mainz\cite{mainz} and E93-049 \cite{strauch} along with preliminary results from E03-104. The data are compared to calculations from the Madrid group\cite{udias} and Schiavilla et al.\cite{schiavil}. The comparison is made for missing momentum $p_m \approx 0$; note that the experimental data have been corrected for the spectrometer acceptance. \label{fig:inducpol}}
\end{minipage}
\end{center}
\end{figure}

The results for the induced polarization $P_y$ are shown in fig. \ref{fig:inducpol}. As mentioned earlier, $P_y$ is directly sensitive to FSI effects. The present large size of the systematic errors is due to the uncertainty in instrumental asymmetries that will decrease strongly with further analysis. The data have been corrected for the HRS acceptance to facilitate a comparison with the calculations. The data suggest that the FSI effects have been overestimated in Schiavilla's calculations. It will be interesting to see how his calculations for $R$ will change once he has optimized his calculations for $P_y$.

\section{The Decay Width of the Neutral Pion}

The two-photon decay mode of the neutral pion $\pi^0$ is completely due to the quantum fluctuations of the quark fields coupling to a gauge field, a symmetry breaking of purely quantum-mechanical origin. In the chiral SU(2) limit of vanishing quark masses its decay width can be expressed in terms of the pion decay constant $F_{\pi}$ and the fine structure constant $\alpha$

\begin{equation}
\Gamma_{\pi^0 \rightarrow \gamma \gamma} =\left[ \frac{M_{\pi^0}}{4\pi}\right]^3\left[\frac{\alpha}{F_{\pi}}\right]^2 = 7.725 \pm 0.044 \rm{~eV}
\end{equation}

Recently, two calculations\cite{nlo, ioffe} have studied the effect of the explicit breaking of the SU(2) symmetry induced by the non-zero quark masses in next-to-leading order, resulting in predictions for the decay width of 8.10 $\pm$ 0.08 eV and 7.93 $\pm$ 0.12 eV. The main contribution to the shift upwards of nearly 0.4 eV is due to isospin breaking that introduces components from the $\eta$ and $\eta ^\prime$ in the physical $\pi^0$.

\begin{figure}[h!]
\begin{center}
\begin{minipage}[t]{8 cm}
\includegraphics[width=1.1\linewidth]{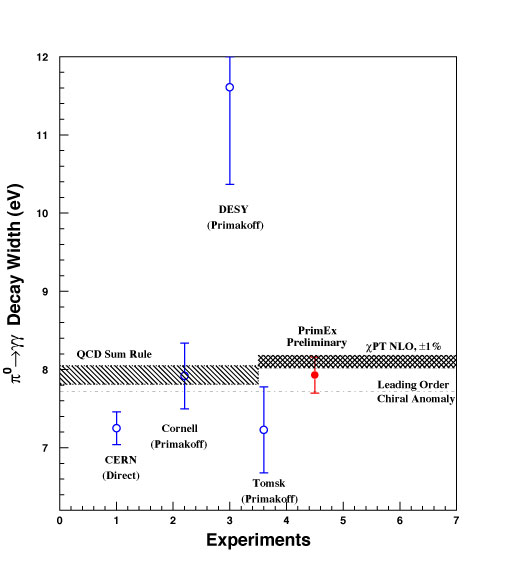}
\end{minipage}
\begin{minipage}[t]{16.5 cm}
\caption{Preliminary result for the $\pi^0 \rightarrow \gamma \gamma$ decay width from the $^{12}$C data of the PRIMEX experiment\cite{e02-103} compared to earlier results from CERN\cite{cern}, Cornell\cite{cornell}, Tomsk\cite{tomsk} and DESY\cite{desy}. The predictions from Chiral Perturbation Theory are also shown, to leading and to next-to-leading order\cite{nlo} and one using the QCD Sum Rule\cite{ioffe}.\label{fig:primexres}}
\end{minipage}
\end{center}
\end{figure}

 However, the experimental knowledge of the decay width is much less established, with a current world-average value of 7.74 $\pm$ 0.55 eV. In the first place the assigned error is nearly an order of magnitude larger than that of the theoretical estimate, but the experimental results show an even larger dispersion. Two methods have been used to measure the $\pi^0$ decay width (or equivalently its lifetime). A direct measurement can be made by observing the decay distance between the production and decay points, using a two-foil technique, in which the pion is produced in the first foil and the decay photons are pair converted in the second foil. The only such direct measurement has been performed at the CERN SPS\cite{cern}. The alternative method uses the Primakoff effect, in which photons produce pions in the Coulomb field of a nucleus. Three such measurements, all using bremsstrahlung beams, have been reported\cite{cornell, tomsk, desy} with a large dispersion as is evident from fig. \ref{fig:primexres}.

\begin{figure}[h!]
\begin{center}
\begin{minipage}[t]{8 cm}
\includegraphics[width=1.2\linewidth]{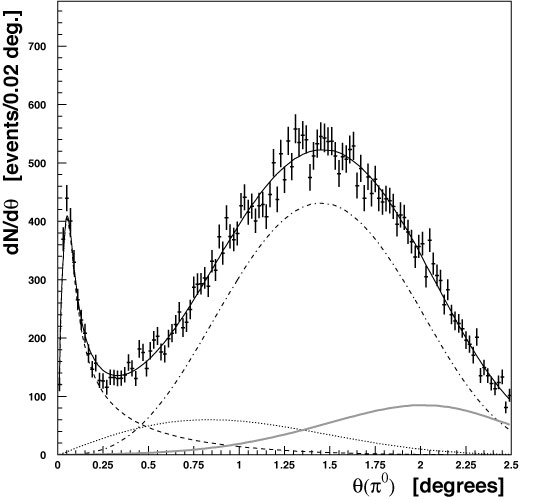}
\end{minipage}
\begin{minipage}[t]{16.5 cm}
\caption{The measured angular dependence of the total $\pi^0$ cross section on $^{12}$C, together with the contributions from the Primakofff effect (dashed), from nuclear coherent (dot-dashed) and incoherent (grey full) scattering and from the interference (short-dashed) between Primakoff and coherent scattering.\label{fig:primexcross}}
\end{minipage}
\end{center}
\end{figure}

The tagged photon facility in Hall B at JLab provides significant advantages for a new high-precision measurement of the $\pi^0$ decay width using the Primakoff effect: the well-defined photon energy enables a clean separation of background processes and the tagging technique allows a better control of systematic errors. Besides the Primakoff effect, three additional processes contribute to pion photoproduction at high energies: nuclear coherent and incoherent scattering and interference between Primakoff and coherent scattering. These processes can be separated by their different angular behaviors (see fig. \ref{fig:primexcross}). Consequently, a detector is needed for the $\pi^0$ detection with good angular resolution to identify the competing processes and good energy resolution to suppress multi-photon backgrounds through cuts on the invariant mass. A hybrid design was chosen for this calorimeter (HyCal) in order to optimize the performance and cost. The central part contains 1152 PbWO$_4$ crystals, 20.5 x 20.5 x 180 mm$^3$ each, arranged in a square array with a four-block hole in the center to allow the beam to pass through. The central part is surrounded by 576 lead glass blocks, 38.2 x 38.2 x 450 mm$^3$ each. The total size of the calorimeter, 116 x 116 cm$^2$, is sufficient to detect both decay photons from a $\pi^0$ meson produced at the target 7.3 m upstream. An essential requirement for the experiment is to determine the cross section with high absolute accuracy. The photon flux was measured with a lead-glass total absorption counter that was put into the beam periodically and monitored continuously with a pair spectrometer that detected the e$^+$e$^-$ pairs produced in the target using a dipole magnet and a set of scintillators. The thickness of the two targets used, $^{12}$C and $^{208}$Pb, was measured to an accuracy of 0.04\%.

The experiment\cite{e02-103} ran for three months (one month for commissioning and two for production running) in late 2004 at a beam energy of 5.7 GeV. The energy and angular resolution of the central part of the calorimeter were shown to be 2.3\% (at a photon energy of 1 GeV) and 0.34 mrad, respectively, resulting in an invariant mass resolution of 2.3 MeV/$c^2$ at the $\pi^0$ mass. A very clean identification of elastic $\pi^0$ events was obtained after a final correlated cut on the invariant mass and the so-called elasticity - the ratio of the energy of a cluster pair and the tagger energy -. The angular distribution of those events, using only the central part of the calorimeter, is shown in fig. \ref{fig:primexcross}. Finally, the Primakoff cross section was extracted from a fit to the angular distribution in which the amplitudes for the Primakoff, coherent and incoherent scattering and the phase of the interference term were varied. The preliminary result for the $\pi^0$ decay width from only the $^{12}$C data is 7.93 eV $\pm$ 2.1\% $\pm$ 2.0\%, in excellent agreement with the NLO predictions. The main limitations to reach the projected goal of a 1.4\% accuracy are statistics (presently 1.8\%, caused by cutting the run short and limiting the analysis so far to the central part of HyCal) and the uncertainty in the photon flux (presently 1.1 \%). In the final analysis both those issues will be addressed.

\section{Electron-Induced Hypernuclear Spectroscopy}

Information about the deeply-lying shell structure of the nuclear interior is difficult to obtain 
since all available nucleon levels are filled, and the addition of a nucleon to a low-lying 
shell is blocked by the Pauli principle.  Because a $\Lambda$ hyperon does not suffer from Pauli blocking, it can be inserted deep inside a nucleus as an impurity, thus providing a sensitive probe of the nuclear interior. In addition, the elementary hyperon-nucleon interaction can be studied with great sensitivity in hypernuclear spectroscopy. Nuclear matter containing strangeness is also predicted to play an important role in stellar objects such as neutron stars. In the past, hypernuclear spectroscopy has been studied with hadronic reactions, severely hampered by the poor beam quality of the secondary pion and kaon beams.

Kaon electro-production from nuclei makes it possible to deposit a tagged $\Lambda$ deep inside 
the nucleus, and to observe its interaction with the nuclear system.  A excellent energy resolution 
(to separate the individual states in the hypernuclear system) and a high beam intensity (to 
compensate for the small cross section) are essential requirements for such studies.  
Electro-production provides a number of advantages over earlier studies\cite{hashirev}:
\begin{itemize}
\item the excellent beam quality makes it possible to identify hypernuclear states with a resolution of potentially $\sim$300 keV 
\item due to the large momentum transferred and strong spin-flip amplitudes  in electro-production hypernuclear states can be created with natural and unnatural parity and low and high spin
\item the ($K^+ - \Lambda$) pair is produced on the proton in contrast to hadronic studies where it is produced on the neutron, making it possible to study different hypernuclei and charge-dependent effects from a comparison of mirror hypernuclei, $e.g.$ $^{12}_\Lambda$B - $^{12}_\Lambda$C
\end{itemize}
The multi-GeV high-intensity beam at JLab thus provides a unique opportunity to study hypernuclear spectroscopy. Two different implementations have been chosen in Halls A and C. 

In Hall C the so-called "zero-degree tagging" configuration was chosen, whereby the production rate was optimized by detecting both the scattered electrons and the kaons at extreme  forward angles with a beam energy close to the reaction threshold, $\sim$1.8 GeV. A splitter magnet was used to separate the electrons and the kaons to different spectrometers. In a first experiment\cite{miyoshi} a missing-mass resolution of 900 keV was obtained in $^{12}_\Lambda$B, but the production rate was severely limited by accidental coincidences from bremsstrahlung. In a follow-up experiment\cite{e01-011} the electron spectrometer, the same Enge split-pole as in the first experiment, was slightly - by  $\sim$8.5\updeg\ - tilted vertically which significantly suppressed both the bremsstrahlung and the M{\o}ller rate. Also the SOS spectrometer (short-orbit) was replaced by a new high-resolution kaon spectrometer (HKS) with two layers of water and three layers of aerogel \v{C}erenkov detectors for particle identification. In this new configuration data were taken in 2005 with a hypernuclear production rate of $\sim$10 per hour. A preliminary analysis has indicated a missing-mass resolution of $\leq$ 500 keV. For the following stage a new electron spectrometer has been designed and constructed for a tenfold increase in production rate and a energy resolution as small as 300 keV. This experiment\cite{e05-115} is expected to run in 2009.

\begin{figure}[h]
\begin{center}
\begin{minipage}[t]{16 cm}
\includegraphics[width=1.0\linewidth]{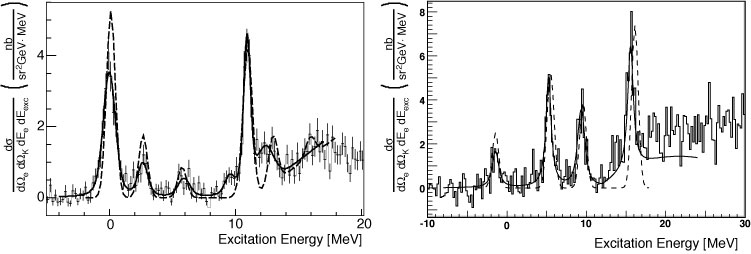}
\end{minipage}
\begin{minipage}[t]{16.5 cm}
\caption{Left: The excitation-energy spectrum of $^{12}_\Lambda$B. The results of the best fit (solid curve) and that of an unnormalized theoretical calculation (dashed curve) are shown superimposed on the data. See text for further details. Right: The same for $^{16}_\Lambda$N. \label{fig:hyper-HallA}}
\end{minipage}
\end{center}
\end{figure}

In Hall A the pair of high-resolution spectrometers (HRS) was used, supplemented with a superconducting septum magnet to permit each HRS to detect particles at angles a small as 6.5\updeg. Highly selective particle identification was obtained through the use of a Ring-Imaging \v{C}erenkov (RICH) detector in a proximity-fusing geometry, a 15 mm freon radiator and a CsI photocathode, coupled to two aerogel \v{C}erenkov detectors. Because the path length in each HRS is $\sim$20 m, kaons had to be produced with a rather large momentum, so a beam energy of 4 GeV was selected.

Experiment E94-107\cite{e94-107} took data on a $^{12}$C target in 2004 and on a waterfall target for hypernuclear production on oxygen in 2005. The measured excitation energy ($E_x$) spectrum for $^{12}_\Lambda$B\cite{iodice} is shown in the left part of fig. \ref{fig:hyper-HallA}. After subtracting the background evaluated from random coincidences in a large timing window, no residual background was observable in the negative $E_x$-range. Also shown in fig. \ref{fig:hyper-HallA} is a fit to the six regions with an excess of counts above background. An energy resolution of better than 700 keV FWHM was obtained for the peaks fitted, in large part due to carefully minimizing the primary energy spread of the electron beam - to less than $6 \cdot 10^{-5}$ FWHM - and stabilizing the beam energy centroid during the experiment. Theoretical calculations in the Distorted Wave Impulse Approximation (DWIA) using the SLA model\cite{sla} for the elementary $p(e,e^\prime K^+)\Lambda$ reaction and shell-model wave functions for $^{11}$B, show a very good overall agreement with the data without any normalization. This is the first time that in a hypernuclear spectrum a measurable strength has been observed in the core-excited part. The structures at $\sim$2.7 and $\sim$5.8 MeV can be well described by coupling a $\Lambda$ in a $s$ state to a 1/2$^-$ and a 3/2$^-$ state in $^{11}$B. In the right part of fig. \ref{fig:hyper-HallA} the hypernuclear spectrum is shown for $^{16}_\Lambda$N\cite{cusanno}. Here, four peak regions could be identified. A similar calculation as for $^{12}_\Lambda$B did reproduce the observed cross section for the four peaks, but with non-negliglible discrepancies in the excitation energies. The use of a waterfall target provided simultaneous cross-section data for the elementary $p(e,e^\prime K^+)\Lambda$ reaction.

\section{Short-Range Nucleon-Nucleon Correlations}

The nucleon-nucleon force is strongly attractive at long range, but at short range the nuclear force changes sign and then becomes highly repulsive. A survey of proton knock-out by electrons\cite{lapikas} has shown that the spectroscopic strengths for valence orbitals amounts to only $\sim$60\% of its sum rule value. Most of this depletion was assigned to nucleon-nucleon short-range and tensor correlations, that have moved the strength to large values of the missing energy $E_m$ and the missing momentum $p_m$. A systematic study\cite{rohe} of the $^{12}$C($e,e^\prime p$) reaction in parallel kinematics established $\sim$11\% correlated proton strength at $p_m$-values above the Fermi surface, but $E_m \leq$ 80 MeV, and $\sim$12\% at $E_m \geq$ 80 MeV. Strongly correlated nucleon pairs are characterized by a large relative momentum, but a small center-of-mass momentum. Indeed, the existence of short-range N-N (and 3N) correlations has been quantitatively established through 
the scaling behavior at large energy and momentum transfer, $\omega$ and $Q^2$, as a function of $x$ of inclusive electron scattering~\cite{egiyan}. 
The ratio of the inclusive cross section of a number of nuclei $vs.$ that of $^3$He was observed to become constant at $x \approx 2$ and again at $x \approx 3$, clearly indicating scattering off 2(3)-body correlated states. From the values of these ratios it was established that $\sim$20\% of all nucleons in $^{12}$C occur in correlated pairs (and $\sim$0.2\% in three-nucleon configurations). Benhar {\it et al.}\cite{benhar} have criticized this interpretation, claiming that FSI effects were not taken into account. However, Laget\cite{laget} and Frankfurt\cite{frank} have shown that FSI effects are indeed very large from the correlated configurations, but then cancel in the ratio. 

\begin{figure}[h]
\begin{center}
\begin{minipage}[t]{16 cm}
\includegraphics[width=1.0\linewidth]{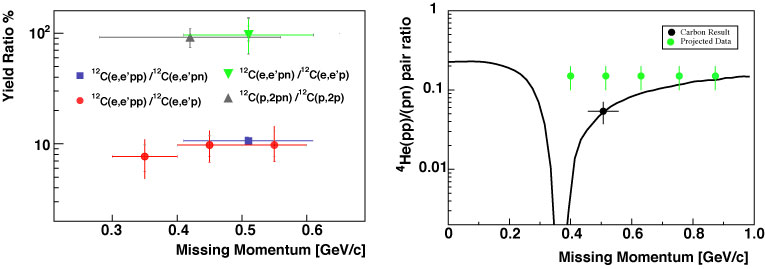}
\end{minipage}
\begin{minipage}[t]{16.5 cm}
\caption{Left: the measured $^{12}$C($e,e'pp$)/$^{12}$C($e,e'pn$) ratio\cite{subedi}, the extrapolated $^{12}$C($e,e'pp$)/$^{12}$C($e,e'p$) and $^{12}$C($e,e'pn$)/$^{12}$C($e,e'p$) yield ratios and the yield ratio of the $^{12}$C($p,2pn$)/$^{12}$C($p,2p$) reaction from ref. (\cite{tang}). Right: the ration of the $pp$ and the $pn$ momentum distributions as calculated by Schiavilla {\it et al.}\cite{schiavil2} compared to the ratio of $pp$ and $pn$ pairs extracted from the measured $^{12}$C($e,e'pp$)/$^{12}$C($e,e'pn$) ratio. Also shown are the projected $p_m$-values and errors for an approved experiment to measure the ratio $^{4}$He($e,e'pp$)/$^{4}$He($e,e'pn$)\cite{e07-006}. \label{fig:srcresult}}
\end{minipage}
\end{center}
\end{figure}

Experiment E01-015\cite{e01-015} studied simultaneously the $^{12}$C($e,e^\prime p$), $^{12}$C($e,e^\prime pp$)  and $^{12}$C($e,e^\prime pn$) reactions. It ran in Hall A in 2004 at a beam energy of 4.6 GeV. The two HRS spectrometers were used to identify the $^{12}$C($e,e^\prime p$) reaction. The kinematic conditions for single-proton knock-out were ($q$, $\omega$, $x$) = (1.65 GeV/$c$, 0.865 GeV, 1.2). Knocked-out protons were detected in the right HRS at three $p_m$-values: 0.35, 0.45 and 0.55 GeV/$c$. E01-015 was the first experiment to use the large acceptance spectrometer BigBite, consisting of a large dipole magnet and equipped with a detector package consisting of three planes of plastic scintillators, segmented in the dispersive direction. Positioned at 1.1 m from the target, it provided an angular acceptance of 96 msr and a momentum acceptance from 0.25 to 0.9 GeV/$c$. Directly behind BigBite, a 0.4 m thick neutron detector was positioned at a distance of 6 m from the target. It contained 88 plastic scintillators, arranged to match the solid angle of BigBite by covering an area of 1 x 3 m$^2$. These unshielded systems operated reliably at a luminosity of $\sim10^{38}$ cm$^{-2}$s$^{-1}$. BigBite was positioned to detect a proton that balanced the $p_m$ for the $^{12}$C($e,e^\prime p$) reaction. Such recoiling protons were identified by requiring an energy loss and a flight time measured with the scintillators  to be consistent with the momentum setting of BigBite. Contributions from $\Delta$-resonance excitation were removed by angular cuts on $\vec{p_m}$. The observed angular distribution of the coincident proton pairs was in agreement with the correlated pair having a CM motion relative to the $A-2$ spectator system with a width of $\sim$140 MeV/c. As shown in fig. \ref{fig:srcresult} for 9.5 $\pm$ 2\% of the $^{12}$($e,e^\prime p$) events, constant in the $p_m$-range from 300 to 600 MeV/$c$, a second proton is ejected roughly back-to-back with the first one\cite{shneor}. Estimates of FSI and charge-exchange contributions were obtained using the Glauber approximation and were shown to cancel each other to a large extent. This conclusion is supported by the isotropy of the angular distribution of the coincidences, mentioned earlier. For each $^{12}$C($e,e^\prime p$) event, both the BigBite and the neutron detector were read out, so that proton and neutron data were collected under identical conditions. The ratio of $^{12}C$($e,e^\prime pn$)/$^{12}C$($e,e^\prime pp$) events was measured to be 9.0 $\pm$ 2.5\cite{subedi}. Since the experiment can only detect 50\% of the initial-state $p-n$ pairs (those with the proton momentum anti-parallel to the virtual photon), the ratio of $p-n$ and $p-p$ pairs is 18 $\pm$ 5. The ratio $^{12}C$($e,e^\prime pn$)/$^{12}C$($e,e^\prime p$) measured to be 0.96 $\pm$ 0.23 (fig. \ref{fig:srcresult}) indicates that in the $p_m$-range of 400 to 600 MeV/$c$ all protons are in a pair configuration.
Recent calculations~\cite{schiavil2,sa05} have shown that the observed dominance of $p-n$ pairs is a clear fingerprint of the short-range tensor component of the nucleon-nucleon force, independent of the parametrization of the nucleon-nucleon force or of the nuclear wave function. In a future experiment~\cite{e07-006} the ratio of ($p-n$)-pairs to ($p-p$)-pairs knocked out from $^4$He will be studied with high statistics over a large $p_m$-range to further investigate the validity of the tensor-correlation  dominance (see right hand part of fig. \ref{fig:srcresult}).

\section{Neutron Charge Form Factor}
\label{GEn}
The nucleon electro-magnetic form factors are of fundamental importance for the understanding of the nucleon's internal structure. 
Early \GEn-experiments used (quasi-)elastic scattering off the deuteron to extract the longitudinal deuteron response function. Due to the smallness of \GEn, the use of
different nucleon-nucleon potentials resulted in a 100\% spread in the resulting 
\GEn. In the past decade a series of double-polarization measurements of
neutron knock-out from a polarized $^2$H or $^3$He target have 
provided accurate data on \GEn. The ratio of the beam-target 
asymmetry with the target polarization perpendicular and 
parallel to the momentum transfer is directly proportional to 
the ratio of the electric and magnetic form factors,

\begin{equation} 
 \frac{\GEn}{\GMn} = - \frac{P_x}{P_z} \frac{\Ee + \Ee'}{2M} \tan(\frac{\thetae}{2}),
\end{equation}
 
\setlength{\parindent}{0em}
where $P_x$ and $P_z$ denote the polarization component perpendicular and parallel to $\vec{q}$.
A similar result is obtained with an unpolarized deuteron target when one 
measures the polarization of the knocked-out neutron as a function of the angle over 
which the neutron spin is precessed with a dipole magnet.

Experiment E02-013\cite{e02-013} measured the charge form factor of the neutron in 2005 by studying the spin asymmetry in the reaction 
$\vec{^3{\rm He}}(\vec e, e^{\prime}n)$ at four values of $Q^2$ up to 3.5 GeV$^2$.  
The scattered electron and the knocked-out neutron were detected in coincidence 
using the open-geometry electron spectrometer 
BigBite and the large neutron detector BigHAND. 

The BigBite spectrometer (with a solid-angle acceptance of roughly 76~msr) was equipped with a detector package containing three wire chambers and a lead-glass shower counter, separated into 
a pre-shower and full-absorption region.  The front and back wire chambers each had six planes of sensitive wires, the middle chamber three planes. 
A hit above a high threshold in the shower counter together with a neutron 
event from BigHAND provided the basic coincidence trigger for the experiment.  

The BigHAND neutron detector (with an active detection area of around $8\,{\rm m}^2$ and a total weight of $\sim$80 tons) contained over 200 neutron bars arranged 
in seven vertical walls sandwiched between iron.  
It also contained two veto walls with $\sim$180 veto counters protected 
from the target by two inches of lead.  

The polarized $^3$He target that sat on the pivot in Hall A was basically designed from the ground up.  
A magnetic holding field of roughly 2 mT was produced using a large iron box 
that also provided shielding from the fringe field of BigBite and served
as a scattering chamber.  
The sealed glass target cells in which the $^3$He was polarized contained 
a mixture of potassium and rubidium, in contrast with previous 
target cells in which the only alkali-metal present was rubidium.  
This hybrid optical pumping technique yielded a substantially higher polarization, quicker pump-up times, and less sensitivity to depolarization from the passage of the electron beam.  Light from high-power diode-laser arrays was brought to the target using optical fibers.  The optics used for polarizing the light  and focusing it onto the target cells were mounted directly on top of the target enclosure.  The target held a polarization of about 50\% during many weeks of
continuous running with an 8~$\mu$A electron beam.
With the 40~cm long target cells containing  $^3$He at roughly 10~atm, E02-013 operated at a luminosity of $5\cdot 10^{36}$~/cm$^2$/s.   
With such a high luminosity, improved target polarization, and the large 
acceptance of the BigBite/BigHAND combination, a 
Figure-of-Merit was achieved that was at least 15 times larger than that of 
any previous $G_E^n$ experiment.    

\begin{figure}[tb]
\begin{center}
\begin{minipage}[t]{16 cm}
\includegraphics[width=1.0\linewidth]{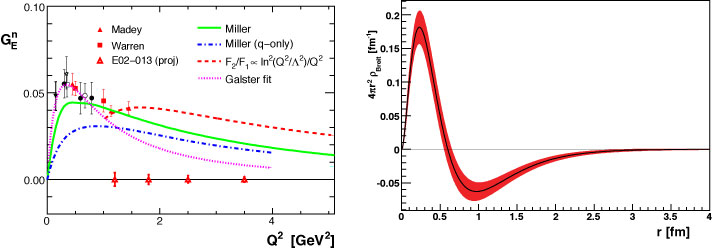}
\end{minipage}
\begin{minipage}[t]{16.5 cm}
\caption{Left: Selected world data on  $G_E^n$ with
the values of $Q^2$ in E02-013 and the expected accuracy, shown relative to the Galster fit.
The solid curve is a calculation in the relativistic constituent quark model 
by G.~Miller~\cite{miller}. The dashed curve takes its shape from the pQCD prediction 
by A.~Belitsky~{\it et al.}\cite{pQCD}, normalized to the experimental $G_E^n$ value at 1.3~GeV$^2$.
Right: The radial distribution of the charge in the neutron, extracted from an analysis of the world data.\label{fig:expected-result}}
\end{minipage}
\end{center}
\end{figure}

The analysis of the BigBite optics resulted in a
vertex reconstruction of 5~mm and a momentum resolution of $\sim$1\%.   
The shower detector 
was calibrated before the experiment with cosmic rays.
A resolution of $\sim$0.5~ns after corrections was achieved for the time of flight (TOF) measurement in BigHAND. 
The veto efficiency per counter is about 95\%, consistent with 
expectations at the high counting rate.
Quasi-elastic events were selected through 
cuts on the invariant mass $W$ and on the transverse component of the missing momentum $P_{per}$. 
The dilution of the measured beam asymmetry due to proton leakage through charge-exchange reactions was extracted from measurements on various targets and found to be in good agreement with simulations. 
A preliminary analysis of the data at Q$^2$ = 1.8 GeV$^2$ has yielded a value of \GEn\ close to the Belitsky scaling prediction with an accuracy of $\sim$15\%, which is the total error expected for each of the four \Q-values, as illustrated 
in the left part of Fig.~\ref{fig:expected-result}.  

In the Breit frame the nucleon form factors can be written as Fourier transforms of the charge
and magnetization distributions. However, if the wavelength of the probe is larger than the Compton wavelength of the nucleon, i.e. if $| Q | \ge M_N$, the form factors also contain dynamical effects due to relativistic boosts. The negative lobe at $\sim$1 fm in the radial charge distribution shown in the right part of Fig.~\ref{fig:expected-result} supports the model of the neutron in which for part of the time it consists of a $\pi^-$ circling a proton, which is consistent with Miller's calculation that shows a large contribution of the pion at low $Q^2$-values. Also, at these $Q^2$-values the relativistic corrections in the Fourier transforms are expected to be small.

\section{Deeply Virtual Compton Scattering}

The Generalized Parton Distributions (GPD) relate the spatial and momentum distributions
of a parton in a nucleon by providing a consistent framework for the EMFF and the parton distribution functions (PDF). Deeply Virtual Compton scattering (DVCS) is the most accessible reaction to study GPDs. 
The E00-110 experiment\cite{e00-110} ran in 2004 with 5.75
GeV electrons  incident on a 15 cm long liquid H$_2$ target. The
luminosity was typically $10^{37}$/cm$^2$/s with a $76\%$ beam polarization. 
The scattered electrons were detected in one High Resolution Spectrometer
(HRS),  while photons above a 1~GeV energy threshold (and $\gamma\gamma$
coincidences from $\pi^0$ decay) were detected in a $11\times 12$ array of
$3 \times 3 \times 18.6$ cm$^3$ PbF$_2$ crystals, whose front face was
located 110 cm from the target center. 
DVCS events were selected from electron-photon coincidences, after subtraction of the $\pi^0$ yield - estimated from two-photon events - and application of a missing-mass cut $M_X^2 < (M+m_\pi)^2$.

\begin{figure}[h]
\begin{center}
\begin{minipage}[t]{8 cm}
\includegraphics[width=1.4\linewidth]{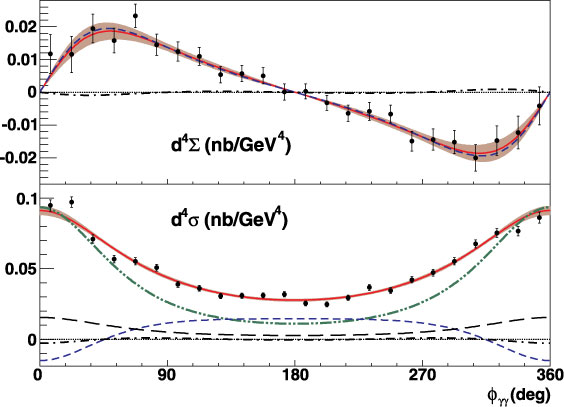}
\end{minipage}
\begin{minipage}[t]{16.5 cm}
\caption{Data\cite{munoz} and fit to $d^4\Sigma$, and $d^4\sigma$, as a function of $\phi_{\gamma\gamma}$, in the bin $\langle Q^2, t\rangle=(2.3,-0.28)$ GeV$^2$ at
$\left\langle x_{\rm Bj} \right\rangle=0.36$. The solid lines show total fits with one-$\sigma$
statistical error bands. The dot-dot-dashed line is the $|{\rm BH}|^2$
contribution to $d^4\sigma$. The short-dashed lines are the contributions from the fitted $\Imag$ and $\Real$
parts of ${\mathcal C}^{\mathcal I}({\mathcal F})$. The
long-dashed line is the fitted $\Real[{\mathcal C}^{\mathcal
I}+\Delta\mathcal C^{\mathcal I}](\mathcal F)$ term.  The dot-dashed
curves are the fitted $\Imag$ and $\Real$ parts of ${\mathcal C}^{\mathcal
I}({\mathcal F}^{\rm eff})$.\label{fig:SigmaDiff}}
\end{minipage}
\end{center}
\end{figure}

To order twist-3 the DVCS helicity-dependent ($d\Sigma$) and
helicity-independent ($d\sigma$) cross sections are given by~\cite{Belitsky}:
\begin{eqnarray}
{d^4\Sigma \over d^4\Phi}  
&\equiv& {1\over 2}
  \left[ {d^4\sigma^+\over d^4\Phi} - {d^4\sigma^-\over d^4\Phi}\right] 
     = \\
&=& \sin(\phi_{\gamma\gamma}) \Gamma_{1}^{\Im} \,
  \Im {\rm m} \left[{\mathcal C^I}({\mathcal F})\right]  
  -  \sin(2\phi_{\gamma\gamma}) 
     \Gamma_{2}^{\Im} \,
    {\Im {\rm m}}\left[{\mathcal C^I}({\mathcal F}^{\rm eff})\right]\,,  
\nonumber
\label{eq:dSigma}
\\
{d^4\sigma\over d^4\Phi}    
&\equiv& {1\over 2}
     \left[{d^4\sigma^+\over d^4\Phi} + {d^4\sigma^-\over d^4\Phi}\right]
        =  {d^4\sigma(|DVCS|^2) \over dQ^2 dx_{\rm Bj} dt d\phi_{\gamma\gamma}} 
+ {d^4\sigma(|BH|^2) \over dQ^2 dx_{\rm Bj} dt d\phi_{\gamma\gamma}}
 \label{eq:dsigma} \\
&+&  \Gamma_{0,\Delta}^{\Re}  {\Real }\left[
              {\mathcal C^I}+\Delta{\mathcal C^I}\right] ({\mathcal F}) 
         + \Gamma_0^{\Re}  {\Real}\left[{\mathcal C^I}({\mathcal F})\right] 
    \nonumber \\
  &-&  
\cos(\phi_{\gamma\gamma})
  \Gamma_{1}^{\Re} {\Real}\left[{\mathcal C^I}({\mathcal F})\right]
  +  \cos(2\phi_{\gamma\gamma}) 
     \Gamma_{2}^{\Re}
    {\Real} \left[{\mathcal C^I}({\mathcal F}^{\rm eff})\right]\,,
\nonumber
\end{eqnarray}
where $\phi_{\gamma\gamma}$ denotes the
azimuthal angle of the detected photon.
The $\Gamma_{n}^{\Re,\Im}$ are kinematic factors with a 
$\phi_{\gamma\gamma}$ dependence that arises from the electron propagators of
the BH amplitude. The ${\mathcal C^I}$ and $\Delta{\mathcal C^I}$
angular harmonics
depend on the interference of the BH amplitude with the set
${\mathcal F}=
\{{\mathcal H,\,\mathcal E,\,\tilde{\mathcal H},\,\tilde{\mathcal E}}\}$
of twist-2 Compton form factors (CFFs) or the related set
${\mathcal F}^{\rm eff}$ of effective twist-3 CFFs:
\begin{eqnarray}
{\mathcal C^I}({\mathcal F}) = F_1{\mathcal H}+
     \xi G_M \tilde{\mathcal H}
 -{t\over 4 M^2} F_2{\mathcal E}\quad\quad\ \ \,\,\,\ \ 
\label{eq:gpds1} \\
\left[{\mathcal C^I}+\Delta{\mathcal C^I}\right]({\mathcal F}) =
           F_1{\mathcal H}-{t\over 4 M^2} F_2{\mathcal E} 
   - \xi^2 G_M 
       \left[{\mathcal H}+{\mathcal E}\right],\  
\label{eq:gpds3}
\end{eqnarray}

in which $F_1$, $F_2$ and $G_M \equiv F_1 + F_2$ denote the elastic form factors.

$d\Sigma$ measures the imaginary part of the BH-DVCS
interference terms and provides direct access to GPDs at $x=\xi$, while $d\sigma$ determines the real part of the BH-DVCS interference
terms and measures the integral of GPDs over its full domain in $x$.

\begin{figure}[b!]
\begin{center}
\begin{minipage}[t]{8 cm}
\includegraphics[width=1.4\linewidth]{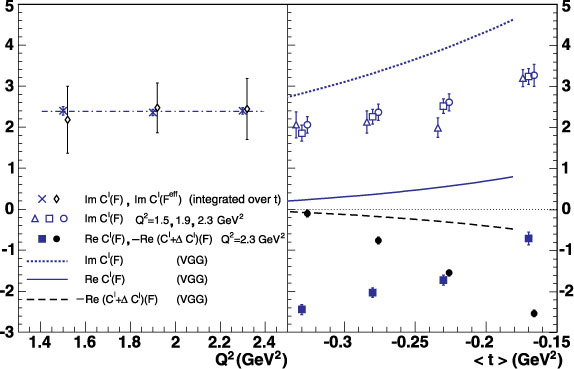}
\end{minipage}
\begin{minipage}[t]{16.5 cm}
\caption{Left:  $Q^2$-dependence of the $\Imag$ parts of (twist-2) 
${\mathcal C}^{\mathcal I}(\mathcal F)$ and (twist-3) 
${\mathcal C}^{\mathcal I}(\mathcal F^{\rm eff})$ 
angular harmonics, averaged over $t$. The
horizontal line is the fitted average of 
$\Imag[{\mathcal C}^{\mathcal I}(\mathcal F)]$.
Right: Extracted real and imaginary parts of the twist-2 and twist-3 angular
harmonics as a function of $t$. 
\label{fig:DVCScoef}}
\end{minipage}
\end{center}
\end{figure}

Figure~\ref{fig:SigmaDiff} shows $d\Sigma$ and $d\sigma$ for one $(Q^2,x_{\rm Bj}, t)$ bin\cite{munoz}.
Clearly, the twist-3 terms make only a very small contribution to the
cross sections. Note also that $d\sigma$ is
much larger than the BH contribution alone, especially from 90$^\circ$ to
270$^\circ$. This indicates that the Beam Spin Asymmetry (BSA =
$d^4\Sigma/d^4\sigma$) can not be simply equated to the imaginary part of the
BH-DVCS interference divided by the BH cross section. 
 Figure~\ref{fig:DVCScoef} (Left) shows the $Q^2$-dependence of the
(twist-2) angular harmonic $\Imag[{\mathcal C}^{\mathcal I}]$ over the
full $t$ domain. 
The absence of a $Q^2$ dependence of
$\Imag[{\mathcal C}^{\mathcal I}(\mathcal F)]$ within its 3\% statistical
uncertainty provides crucial support to the dominance of twist-2 in the
DVCS amplitude. 
$\Imag[{\mathcal C}^{\mathcal I}(\mathcal F)]$ is thus a direct
measurement of the linear combination
of GPDs. 
Figure~\ref{fig:DVCScoef} (Right) displays the
twist-2 ${\mathcal C}$ angular harmonics  as a function of $t$, together with the 
VGG model estimates~\cite{Vanderhaeghen}. 
The VGG model  is in qualitative
agreement with the $\Imag[\mathcal C^{\mathcal I}(\mathcal F)]$ data, but
significantly under-predicts the $\Real$ parts
of the angular harmonics.

\begin{figure}[t!]
\begin{center}
\begin{minipage}[t]{16 cm}
\includegraphics[width=1.0\linewidth]{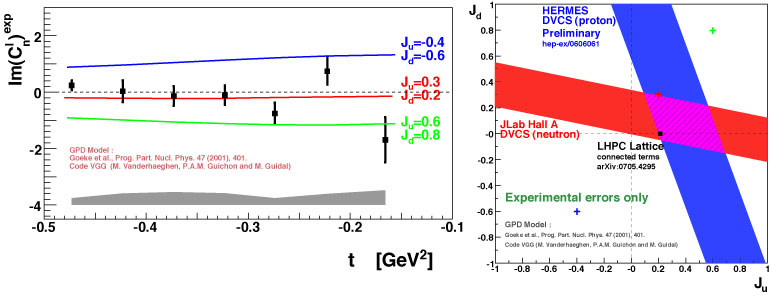}
\end{minipage}
\begin{minipage}[t]{16.5 cm}
\caption{Left: the $t$-dependence of the extracted $\sin (\phi_{\gamma \gamma})$ moments for incoherent n-DVCS, compared to VGG calculations for different values of the orbital angular momenta of the valence quarks, $J_u$ and $J_d$; right: experimental constraint on $J_u$ and $J_d$ from the present n-DVCS results\cite{mazouz}. A similar constraint from HERMES\cite{hermes} and the result from a LQCD-based calculation\cite{LQCD} are also shown.\label{fig:neut}}
\end{minipage}
\end{center}
\end{figure}

Following the E00-110 experiment, the E03-106 experiment\cite{e03-106} 
explored DVCS off the neutron. 
Because of the very small magnitude of the Dirac form factor $F_1$ in the neutron
case, this observable is supposed to be sensitive to $\mathcal E$, the least constrained GPD.
Within the Impulse Approximation the main contributions to electroproduction of photons on the deuteron come from coherent (d-DVCS) and incoherent (p-DVCS) and (n-DVCS) scattering.
The n-DVCS cross-section difference is obtained by first subtracting the proton contribution measured on a hydrogen target and then
separating the neutron and deuteron contributions in each $t$-bin via a global analysis,
that used the dynamical separation ($-t/2$) between 
these two channels and the different kinematical factors that dominate 
their $\sin(\phi)$ moments. 
The resulting neutron moments turn out to be small and 
negative (Fig.~\ref{fig:neut}), consistent with theoretical expectations; the 
sensivity to the GPD $\mathcal E$ through the quark angular momentum is also shown on 
the figure. A correlated constraint on the orbital angular momenta of the valence quarks, $J_u$ and $J_d$, is then extracted\cite{mazouz} from a fit with the VGG model\cite{Vanderhaeghen}.

\section{Summary}

Mainly preliminary results were presented for eight recent JLab experiments. The BONUS detector has been successfully commissioned and will provide detailed information on the neutron structure, unbiased by nuclear corrections. Polarization transfer in $^4$He has shown to be a sensitive test of possible modifications of the nucleon form factors by the nuclear medium. A new study of the Primakoff effect has yielded accurate information on the $\pi ^0$ decay width. A new class of hypernuclear spectroscopy studies with much improved energy resolution has been initiated thanks to the excellent beam quality at JLab. Triple coincident nucleon knock-out studies have yielded quantitative information on short-range nucleon-nucleon correlations. The analysis of experiment E02-013 will extend the existing data on \GEn\ to double the $Q^2$-range. Experiment E00-110 has yielded  the first measurements of the DVCS cross
section in the valence quark region. The $Q^2$-dependence of the angular
harmonics of the helicity-dependent cross section provided solid
evidence of twist-2 dominance in DVCS. 
The unexpectedly large contribution of the DVCS$^2$ term observed in the cross section impedes the direct extraction of GPDs from BSA measurements.

\section{Acknowledgements}
The author gratefully acknowledges detailed discussions with drs. H. Fenker, A. Gasparian, J. LeRose, D. Higinbotham, B. Wojtsekhowski and C. Hyde. The permission by the spokespersons of the experiments selected to present preliminary results is greatly appreciated. This work was supported by DOE contract DE-AC05-06OR23177, under which Jefferson Science Associates, LLC, operates the Thomas Jefferson National Accelerator Facility. 

\end{document}